\documentclass[prl,floats,aps,twocolumn,epsf,graphicx]{revtex4}
\usepackage{epsfig}
\usepackage{graphicx}

\begin{document}

\title{Hairy Black Holes and Null Circular Geodesics}
\author{Shahar Hod}
\address{The Ruppin Academic Center, Emeq Hefer 40250, Israel}
\address{}
\address{The Hadassah Institute, Jerusalem 91010, Israel}
\date{\today}

\begin{abstract}

\ \ \ Einstein-matter theories in which hairy black-hole
configurations have been found are studied. We prove that the
non-trivial behavior of the hair must extend {\it beyond} the null
circular orbit (the ``photonsphere") of the corresponding spacetime.
We further conjecture that the region above the photonsphere
contains at least $50\%$ of the total hair's mass. We support this
conjecture with analytical and numerical results.
\end{abstract}
\bigskip
\maketitle

The influential `no-hair conjecture' of Wheeler \cite{Whee} has
played a key role in the development of black-hole physics
\cite{BekTod,Nun}. This conjecture suggests that black holes are
fundamental objects in general relativity, Einstein's theory of
gravity -- they should be described by only a few parameters, very
much like atoms in quantum mechanics.

The no-hair conjecture was motivated by earlier uniqueness theorems
on black-hole solutions of the Einstein vacuum theory and the
Einstein-Maxwell theory \cite{un1,un2,un3,un4,un5}. According to
these uniqueness theorems, all stationary solutions of the
Einstein-Maxwell equations are uniquely described by only three
conserved parameters which are associated with a Gauss-like law:
mass, charge, and angular momentum.

The belief in the no-hair conjecture was based on a simple physical
picture according to which all matter fields left in the exterior of
a newly born black hole would eventually be radiated away to
infinity or be swallowed by the black hole itself (except when those
fields were associated with conserved charges). In accord with this
logic, early no-hair theorems indeed excluded scalar \cite{Chas},
massive vector \cite{BekVec}, and spinor \cite{Hart} fields from the
exterior of stationary black holes.

However, the interplay between particle physics and general
relativity in the following years has led to the somewhat surprising
discovery of various types of ``hairy'' black holes, the first of
which were the ``colored black holes'' \cite{BizCol}. These are
black-hole solutions of the Einstein-Yang-Mills (EYM) theory that
require for their complete specification not only the value of the
mass but also an additional integer, $n$, which counts the number of
nodes of the Yang-Mills field outside the horizon. Remarkably, this
integer is not associated with any conserved charge.

Soon after this discovery, a variety of hairy black-hole solutions
equipped with different types of exterior fields have been
discovered
\cite{Lavr,BizCham,Green,Stra,BiWa,EYMH,Volkov,BiCh,Lav1,Lav2,Bizw,Bek25}.
These include the Einstein-Skyrme, Einstein-non Abelian-Proca,
Einstein-Yang-Mills-Higgs, and Einstein-Yang-Mills-Dilaton hairy
black holes.

It has become clear \cite{Nun} that the {\it non}-linear character
of the matter fields mentioned above plays a key role in the
construction of these hairy black-hole configurations. N\'u\~nez et.
al. \cite{Nun} have presented a nice heuristic picture according to
which it is the self-interaction between the part of the field in a
region {\it near} the black-hole horizon (a loosely defined region
from which the hair tends to be swallowed by the black hole) and the
part of the field in a region relatively {\it far} from the black
hole (a region from which the hair tends to be radiated away to
infinity) which is responsible, together with gravity, for the
existence of stationary black-hole solutions with exterior matter
fields (hair). The non-linear (self-interaction) character of the
fields thus plays an essential role in binding together the hair in
these two regions in such a way that the ``near-horizon" hair does
not collapse into the black hole while the ``far-region" hair does
not escape to infinity.

Thus, according to the heuristic picture of \cite{Nun}, the
non-trivial (non-linear) behavior of the matter fields which
constitute the hair is expected to extend into some loosely defined
``far region" well above the black-hole horizon.
But is it possible to provide a more explicit characterization of
the hair's length?

Here we turn our attention to another important characteristic of
black-hole spacetimes: null geodesics. Geodesic motions provide
important information on the structure of the spacetime geometry.
Among the different kinds of geodesic motion, circular geodesics are
especially important \cite{Chan,CarC}. In particular, the null
circular orbit (also known as the ``photon orbit" or
``photonsphere") is the boundary between two qualitatively different
regions in the exterior of a black hole: No stationary
spherically-symmetric configurations made of test particles (with no
self-interactions) can exists below this orbit \cite{Notephoton}.
Gravity is simply too strong there. Relating this property of the
null circular geodesic to our former discussion on hairy black
holes, we conjecture that the ``near region" (the region from which
the hair tends to be sucked into the black hole) extends at least up
to the height of the photon orbit.

The aim of this Letter is to prove a theorem which supports this
conjecture. We shall show that the non-trivial behavior of the hair
indeed extends into the region {\it above} the photonsphere. More
explicitly, we shall show that the asymptotic behavior of the
exterior fields cannot start before the null circular orbit is
crossed.

The first clue for the important role played by the photonsphere in
determining the effective length of the hair could have been
elicited from the nice theorem proved in \cite{Nun}. There it was
shown that the non-trivial behavior of some suitably defined
pressure function ${\cal E}(r)\equiv e^{-\delta} r^4p$ (see details
below) must extend beyond $3/2$ the horizon radius. But what is so
special in the location $r={3\over 2}r_H$?

Here we point out that the null circular orbit of the (bare)
Schwarzschild spacetime is actually located exactly at ${3\over
2}r_H$. This location is not expected to change much for hairy black
holes with ``thin" hair (these are characterized by
$m_{\text{hair}}\ll M$, where $M$ is the total ADM mass of the
spacetime and $m_{\text{hair}}\equiv M-{1\over 2}r_H$ is the mass of
the hair which resides outside the horizon). This suggests that, for
black holes with thin hair, the non-trivial (non-asymptotic)
behavior of the fields extends beyond the photonsphere. Here we
shall show that this is actually a generic property of hairy
black-hole configurations, regardless of the amount of hair.

We consider static spherically symmetric asymptotically flat
spacetimes. The line element may take the following form in
Schwarzschild coordinates \cite{Nun,Hodm}
\begin{equation}\label{Eq1}
ds^2=-e^{-2\delta}\mu dt^2 +\mu^{-1}dr^2+r^2(d\theta^2 +\sin^2\theta
d\phi^2)\  ,
\end{equation}
where the metric functions $\delta(r)$ and $\mu(r)\equiv 1-2m(r)/r$
depend only on the Schwarzschild areal coordinate $r$. Asymptotic
flatness requires that as $r \to \infty$,
\begin{equation}\label{Eq2}
\mu(r) \to 1\ \ \ {\text{and}}\ \ \ \delta(r) \to 0\  ,
\end{equation}
and a regular event horizon at $r=r_H$ requires \cite{Nun}
\begin{equation}\label{Eq3}
\mu(r_H) =0\ \ \ {\text{and}}\ \ \ \delta(r_H)<\infty\  .
\end{equation}

Taking $T^{t}_{t}=-\rho$, $T^{r}_{r}=p$, and
$T^{\theta}_{\theta}=T^{\phi}_{\phi}=p_T$, where $\rho$, $p$, and
$p_T$ are identified as the energy density, radial pressure, and
tangential pressure respectively \cite{Bond1}, the Einstein
equations $G^{\mu}_{\nu}=8\pi T^{\mu}_{\nu}$ read
\begin{equation}\label{Eq4}
\mu'=-8\pi r\rho+(1-\mu)/r\  ,
\end{equation}
and
\begin{equation}\label{Eq5}
\delta'=-4\pi r(\rho +p)/\mu\  ,
\end{equation}
where the prime stands for differentiation with respect to $r$. (We
use natural units in which $G=c=1$.)

The mass $m(r)$ contained within a sphere of radius $r$ is given by
\begin{equation}\label{Eq6}
m(r)={1\over2}r_H+\int_{r_H}^{r} 4\pi r'^{2} \rho(r')dr'\  ,
\end{equation}
where $m(r_H)=r_H/2$ is the horizon mass.

The conservation equation, $T^{\mu}_{\nu ;\mu}=0$, has only one
nontrivial component \cite{Nun}
\begin{equation}\label{Eq7}
T^{\mu}_{r ;\mu}=0\  .
\end{equation}
Substituting Eqs. (\ref{Eq4}) and (\ref{Eq5}) in Eq. (\ref{Eq7}),
one finds for the pressure gradient
\begin{eqnarray}\label{Eq8}
p'(r)&=& {{1} \over {2\mu r}}\Big[(3\mu-1-8\pi
r^2p)(\rho+p)+2\mu T\nonumber \\
&& -8\mu p\Big]\ ,
\end{eqnarray}
where $T=-\rho+p+2p_T$ is the trace of the energy momentum tensor.
Below we shall analyze the behavior of the function $P(r) \equiv
r^{4}p(r)$, whose derivative is given by
\begin{eqnarray}\label{Eq9}
P'(r)&=& {{r^{3}} \over {2\mu}}\Big[(3\mu-1-8\pi r^2p)(\rho+p)+2\mu
T\Big]\ .
\end{eqnarray}

When analyzing the coupled Einstein-matter system, one usually
imposes some energy conditions on the matter fields. We shall assume
that the hair outside the horizon satisfies the following
conditions:
\newline
(1) The weak energy condition (WEC). This means that the energy
density, $\rho$, is positive semidefinite and that it bounds the
pressures. In particular, $|p| \leq \rho$. This implies the
inequality
\begin{equation}\label{Eq10}
\rho+p \geq 0\  .
\end{equation}
(2) The trace of the energy-momentum tensor plays a central role in
determining the spacetime geometry of static configurations
\cite{Bond1}. It is usually assumed to satisfy the relation $p+2p_T
\leq \rho$ (see \cite{Bond1} and references therein), which implies
\begin{equation}\label{Eq11}
T\leq 0\  .
\end{equation}
(3) The energy density $\rho$ goes to zero faster than $r^{-4}$.
This requirement is the natural way to impose the condition that
there are no extra conserved charges (besides the ADM mass) defined
at asymptotic infinity associated with the matter fields \cite{Nun}.
(We recall that the charges defined at spatial infinity, like the
electric charge of the Reissner-Nordstr\"om solution in
Einstein-Maxwell theory, are associated with the $\rho \sim r^{-4}$
asymptotic behavior.) We therefore have the boundary condition
\begin{equation}\label{Eq12}
P(r\to\infty)\to 0\  .
\end{equation}
It should be emphasized that in all Einstein-matter theories in
which hair has been found, these conditions are indeed satisfied
(see details in \cite{Nun}).

We shall next examine the behavior of the function $P(r)$ in the
vicinity of the black-hole horizon. Regularity of the horizon
imposes the requirement (see \cite{Bekreg,Nun} for details):
\begin{equation}\label{Eq13}
-p(r_H)=\rho(r_H)<(8\pi r^2_H)^{-1}\  ,
\end{equation}
the last inequality being valid for non-extremal black holes.
Substituting Eqs. (\ref{Eq3}), (\ref{Eq10}), (\ref{Eq11}) and
(\ref{Eq13}) into Eq. (\ref{Eq9}), one finds
\begin{equation}\label{Eq14}
P(r\to r_H)\leq 0\ \ \ {\text{and}}\ \ \ P'(r\to r_H)<0\
\end{equation}
in the vicinity of the black-hole horizon.

We shall now prove that the asymptotic behavior of the pressure
function $P(r)$, as characterized by Eq. (\ref{Eq12}), can start
only {\it above} the photonsphere. We shall follow the analysis of
\cite{Chan,CarC} in order to compute the location $r=r_{\gamma}$ of
the null circular geodesic for a black-hole spacetime described by
the line element (\ref{Eq1}). The Lagrangian describing the
geodesics in the spacetime (\ref{Eq1}) is given by
\begin{equation}\label{Eq15}
2{\cal L}=-e^{-2\delta}\mu\dot t^2+\mu^{-1}\dot r^2+r^2\dot\phi^2\
,
\end{equation}
where a dot denotes a derivative with respect to proper time. The
generalized momenta derived from this Lagrangian are given by
\cite{Chan,CarC}
\begin{equation}\label{Eq16}
p_t=-e^{-2\delta}\mu\dot t\equiv -E={\text{const}}\  ,
\end{equation}
\begin{equation}\label{Eq17}
p_{\phi}=r^2\dot\phi\equiv L={\text{const}}\  ,
\end{equation}
and
\begin{equation}\label{Eq18}
p_r=\mu^{-1}\dot r\  .
\end{equation}
The Lagrangian is independent of both $t$ and $\phi$. This implies
that $E$ and $L$ are constants of the motion. The Hamiltonian of the
system is given by \cite{Chan,CarC} ${\cal H}=p_t\dot t +p_r\dot r
+p_{\phi}\dot\phi-{\cal L}$, which implies
\begin{eqnarray}\label{Eq19}
2{\cal H}=-E\dot t+L\dot\phi+\mu^{-1}\dot
r^2=\epsilon={\text{const}}\ ,
\end{eqnarray}
where $\epsilon=0$ for null geodesics and $\epsilon=1$ for timelike
geodesics. Substituting Eqs. (\ref{Eq16})-(\ref{Eq17}) into
(\ref{Eq19}), one finds
\begin{equation}\label{Eq20}
\dot
r^2=\mu\Big[{{E^2}\over{e^{-2\delta}\mu}}-{{L^2}\over{r^2}}-\epsilon\Big]\
.
\end{equation}

Circular geodesics are characterized by $\dot r^2=(\dot r^2)^{'}=0$
\cite{Chan,CarC}. This implies the relations
\begin{equation}\label{Eq21}
E^2={{2e^{-4\delta}\mu^2}\over{2e^{-2\delta}\mu-r(e^{-2\delta}\mu)^{'}}}\
\ \ ;\ \ \
L^2={{r^3(e^{-2\delta}\mu)^{'}}\over{2e^{-2\delta}\mu-r(e^{-2\delta}\mu)^{'}}}\
,
\end{equation}
for timelike geodesics. The requirement that the energy $E$ be real
enforces the inequality
\begin{equation}\label{Eq22}
2e^{-2\delta}\mu-r(e^{-2\delta}\mu)^{'}>0\  .
\end{equation}

From (\ref{Eq20}) one finds that the radius $r=r_{\gamma}$ of the
null circular geodesic satisfies the relation
\begin{equation}\label{Eq23}
r_{\gamma}={{2e^{-2\delta}\mu}\over {(e^{-2\delta}\mu)^{'}}}\  .
\end{equation}
Substituting the Einstein equations (\ref{Eq4})-(\ref{Eq5}) into
Eqs. (\ref{Eq22})-(\ref{Eq23}), one finds
\begin{equation}\label{Eq24}
3\mu-1-8\pi r^2p\geq 0
\end{equation}
in the spacetime region where circular geodesics are allowed to
exist. The equality sign corresponds to the limiting case of the
null circular geodesic. In case there are several such zeroes, the
photonsphere corresponds to the innermost one \cite{Notephot}.

Finally, substituting Eq. (\ref{Eq24}) into (\ref{Eq9}), one finds
the surprisingly simple relation
\begin{equation}\label{Eq25}
P'(r_{\gamma})=r^3_{\gamma}T\leq0\  ,
\end{equation}
where the last inequality follows from (\ref{Eq11}). The spacetime
region between the horizon and the photonsphere in which circular
geodesics are excluded (we refer to this region as the ``no-circling
zone") is characterized by $3\mu-1-8\pi r^2p<0$. This implies
\begin{equation}\label{Eq26}
P'(r<r_{\gamma})\leq0\  .
\end{equation}
Thus, Eqs. (\ref{Eq14}) and (\ref{Eq26}) imply that $P(r)$ is a
non-positive and decreasing function at least up to the point where
the photonsphere is crossed. If we define $r=r_{\text{hair}}$ to be
the point at which $|P(r)|$ has a local maximum [Eqs. (\ref{Eq12})
and (\ref{Eq14}) together imply that such a point must exist and
that it must be crossed before the trivial asymptotic behavior
(\ref{Eq12}) dominates], then our analysis reveals the lower bound
\begin{equation}\label{Eq27}
r_{\text{hair}}\geq r_{\gamma}\  .
\end{equation}
Thus, the nontrivial ({\it non}-asymptotic) behavior of the hair
must extend {\it beyond} the photonsphere \cite{Notemax}. Note that
our pressure function $P(r)\equiv r^4p$ is different from the
function ${\cal E}(r)\equiv e^{-\delta} r^4p$ considered in
\cite{Nun}. Thus, the definitions of the length of the hair adopted
in these papers are different from each other.

An interesting quantity which characterizes the spatial distribution
of the hair is given by the dimensionless ratio
$m^{+}_{\text{hair}}/m^{-}_{\text{hair}}$, where
\begin{equation}\label{Eq28}
m^{+}_{\text{hair}}\equiv M-m(r_{\gamma})
\end{equation}
is the mass of the hair which resides above the photonsphere, and
\begin{equation}\label{Eq29}
m^{-}_{\text{hair}}\equiv m(r_{\gamma})-m(r_H)
\end{equation}
is the mass of the hair which is contained between the horizon and
the photonsphere. (Here $M$ is the total ADM mass of the spacetime.)
The result (\ref{Eq27}) suggests (but obviously does not prove) that
a considerable fraction of the hair's mass resides {\it above} the
photonsphere. This raises the following question: Is there some
fundamental lower bound on the ratio
$m^{+}_{\text{hair}}/m^{-}_{\text{hair}}$ for hairy black holes?

To answer this interesting question, we shall first examine the
limiting case of the {\it linear} Maxwell field outside the
Reissner-Nordstr\"om (RN) black-hole solution. Of course, this is
{\it not} a case where a genuine hair is present since an additional
conserved charge is needed in order to complete the specification of
the solution. Nevertheless, for this black-hole solution (like in
any real hairy solution) the region exterior to the horizon is
characterized by a non-zero energy density. For the Maxwell field
one has $-p(r)=\rho(r)=Q^2/8\pi r^4$ \cite{Bekreg}, which yields
$m(r)=M-Q^2/2r$ for the mass function and $r_{\gamma}={1\over
2}[3M+(9M^2-8Q^2)^{1/2}]$ for the location of the photonsphere, see
Eqs. (\ref{Eq6}) and (\ref{Eq24}), respectively. Here $M$ and $Q$
are the total mass and electric charge of the spacetime,
respectively. Substituting these relations into Eqs. (\ref{Eq28})
and (\ref{Eq29}) with $r_H=M+(M^2-Q^2)^{1/2}$, one finds the ratio
\begin{equation}\label{Eq30}
m^{+}_{RN}/m^{-}_{RN}={{1}\over{r_{\gamma}/r_H-1}}\geq 1\ .
\end{equation}
(The case $m^{+}_{RN}/m^{-}_{RN}=1$ corresponds to the extremal
black-hole solution with $Q=M$.)

The result (\ref{Eq30}) for the marginal case of a {\it linear}
``hair" leads us to conjecture that genuine hairy black holes always
satisfy the lower bound
\begin{equation}\label{Eq31}
m^{+}_{\text{hair}}/m^{-}_{\text{hair}}\geq 1\ .
\end{equation}
In other words, the region above the photonsphere always contain at
least $50\%$ of the total hair's mass.

There is one family of hairy black-hole configurations for which the
suggested bound (\ref{Eq31}) can be tested analytically: the
Einstein-Yang-Mills hairy black-hole solutions. The EYM equations
can be solved analytically in the limit of large black holes (for
which $m_{\text{hair}}\ll r_H$), see \cite{Volkov} for details. The
hair of the $n=1$ solution is then described by Eq. (4.8) of
\cite{Volkov}, which yields the ratio
$m^{+}_{\text{hair}}/m^{-}_{\text{hair}}=2.08$. Thus, large EYM
black holes indeed respect the conjectured bound (\ref{Eq31}).

We have also performed some numerical studies in order to put the
conjectured bound (\ref{Eq31}) into test. The models we have
considered include the Einstein-Yang-Mills, Einstein-Skyrme,
Einstein-non Abelian-Proca, Einstein-Yang-Mills-Higgs, and
Einstein-Yang-Mills-Dilaton systems. We have found that all these
hairy black-hole solutions indeed {\it conform} to the suggested
bound (\ref{Eq31}). These studies will be reported elsewhere
\cite{Hodto}.

In summary, in this Letter we have analyzed the non-trivial spatial
behavior of the matter fields outside hairy black holes. In
particular, we have proved a theorem which reveals the important
role played by the null circular geodesic (the photonsphere) in the
context of hairy black-hole configurations. According to this
theorem, the non-trivial structure of the hair must extend {\it
above} the photonsphere of the corresponding spacetime.

Furthermore, motivated by this theorem we have put forward a
conjecture according to which the region above the null circular
geodesic contains at least $50\%$ of the total hair's mass. This
conjecture is supported by numerical computations for a variety of
hairy black-hole configurations \cite{Hodto}.

\bigskip
\noindent {\bf ACKNOWLEDGMENTS}

This research is supported by the Meltzer Science Foundation. I
thank Yael Oren, Arbel M. Ongo and Ayelet B. Lata for stimulating
discussions.


\begin{thebibliography}{99}

\bibitem{Whee} R. Ruffini and J. A. Wheeler, Phys. Today {\bf 24}, 30 (1971);
B. Carter, in {\it Black Holes}, Proceedings of 1972 Session of Ecole d'ete de Physique Theorique,
edited by C. De Witt and B. S. De Witt (Gordon and Breach, New York, 1973).

\bibitem{BekTod} J. D. Bekenstein, Phys. Rev. D {\bf 7}, 2333 (1973); Phys. Today {\bf 33}, 24 (1980);
Phys. Rev. D {\bf 51}, R6608 (1995).

\bibitem{Nun} D. N\'u\~nez, H. Quevedo, and D. Sudarsky, Phys. Rev. Lett. {\bf 76}, 571 (1996).

\bibitem{un1} W. Israel, Phys. Rev. {\bf 164}, 1776 (1967); Commun.
Math. Phys. {\bf 8}, 245 (1968).

\bibitem{un2} B. Carter, Phys. Rev. Lett. {\bf 26}, 331 (1971).

\bibitem{un3} S. W. Hawking, Commun. Math. Phys. {\bf 25}, 152 (1972).

\bibitem{un4} D. C. Robinson, Phys. Rev. D {\bf 10}, 458 (1974); Phys. Rev.
Lett. {\bf 34}, 905 (1975).

\bibitem{un5} J. Isper, Phys. Rev. Lett. {\bf 27}, 529 (1971).

\bibitem{Chas} J. E. Chase, Commun. Math. Phys. {\bf 19}, 276 (1970); J. D. Bekenstein,
Phys. Rev. Lett. {\bf 28}, 452 (1972); C. Teitelboim, Lett. Nuovo Cimento {\bf 3}, 326 (1972).

\bibitem{BekVec} J. D. Bekenstein, Phys. Rev. D {\bf 5}, 1239 (1972); {\bf 5}, 2403 (1972);
M. Heusler, J. Math. Phys. {\bf 33}, 3497 (1992);
D. Sudarsky, Class. Quantum Grav. {\bf 12}, 579 (1995).

\bibitem{Hart} J. Hartle, Phys. Rev. D {\bf 3}, 2938 (1971); C. Teitelboim, Lett.
Nuovo Cimento {\bf 3}, 397 (1972).

\bibitem{BizCol} P. Bizo\'n, Phys. Rev. Lett {\bf 64}, 2844 (1990); M. S. Volkov and
D. V. Gal'tsov, Sov. J. Nucl. Phys. {\bf 51}, 1171 (1990); H. P.
Kuenzle and A. K. M. Masood- ul- Alam, J. Math. Phys. {\bf 31}, 928
(1990).

\bibitem{Lavr} G. Lavrelashvili and D. Maison, Nucl. Phys. B {\bf
410}, 407 (1993).

\bibitem{BizCham} P. Bizo\'n and T. Chamj, Phys. Lett B {\bf 297}, 55 (1992);
M. Heusler, S. Droz, and N. Straumann, Phys. Lett. B {\bf 268}, 371 (1991); {\bf 271}, 61 (1991);
{\bf 258}, 21 (1992).

\bibitem{Green} B. R. Greene, S. D. Mathur, and C. . O'Neill, Phys. Rev. D {\bf 47}, 2242 (1993);
T. Torii, K. Maeda, and T. Tachizawa, Phys. Rev. D {\bf 51}, 1510 (1995).

\bibitem{Stra} N. Straumann and Z. -H Zhou, Phys. Lett. B {\bf 243}, 33 (1990).

\bibitem{BiWa} P. Bizo\'n and R. M. Wald, Phys. Lett. B {\bf 267}, 173 (1991).

\bibitem{EYMH}  N. E. Mavromatos and E. Winstanley, Phys. Rev. D {\bf 53}, 3190 (1996).

\bibitem{Volkov} M. S. Volkov and D. V. Gal\`tsov, Phys. Rept. {\bf 319}, 1 (1999).

\bibitem{BiCh} P. Bizo\'n and T. Chmaj, Phys. Rev. D {\bf 61}, 067501 (2000).

\bibitem{Lav1} G. V. Lavrelashvili and D. Maison, Phys. Lett. B {\bf 343}, 214 (1995).

\bibitem{Lav2} M. S. Volkov, O. Brodbeck, G. V. Lavrelashvili and N.
Straumann, Phys. Lett. B {\bf 349}, 438 (1995).

\bibitem{Bizw} P. Bizo\'n, Phys. Lett B {\bf 259}, 53 (1991).

\bibitem{Bek25} J. D. Bekenstein, arXiv:gr-qc/9605059.


\bibitem{Chan} S. Chandrasekhar, {\it The Mathematical Theory of Black
Holes}, (Oxford University Press, New York, 1983).

\bibitem{CarC}  V. Cardoso, A. S. Miranda, E. Berti, H. Witek and V. T.
Zanchin, Phys. Rev. D {\bf 79}, 064016 (2009).

\bibitem{Notephoton} The circular null geodesic is actually the limiting case
of circular timelike geodesics (that is, the innermost circular
orbit \cite{Chan,CarC}), see details below.

\bibitem{Hodm} S. Hod, Phys. Lett. B {\bf 657}, 255 (2007) [arXiv:0711.4541];
S. Hod, Class. Quant. Grav. {\bf 24}, 6019 (2007) [arXiv:0712.1988];
S. Hod, Phys. Lett. B {\bf 661}, 175 (2008) [arXiv:0803.0608].

\bibitem{Bond1} H. Bondi, Mon. Not. Roy. Astr. Soc. {\bf 259}, 365
(1992).


\bibitem{Bekreg} A. E. Mayo and J. D. Bekenstein, Phys. Rev. D {\bf 54}, 5059 (1996);
N. E. Mavromatos, arXiv:gr-qc/9606008.

\bibitem{Notephot} Note that Eqs. (\ref{Eq3}) and (\ref{Eq13})
implies $3\mu-1-8\pi r^2p<0$ at the black-hole horizon. On the other
hand, Eqs. (\ref{Eq2}) and (\ref{Eq12}) implies $3\mu-1-8\pi r^2p\to
2$ as $r\to\infty$. Thus, there must be some intermediate point at
which $3\mu-1-8\pi r^2p=0$. This corresponds to the location of the
null circular geodesic.

\bibitem{Notemax} We emphasize that our analysis reveals that $P(r)$ is a
non-positive and decreasing function in the interval
$[r_H,r_{\text{hair}}]$, with $r_{\text{hair}}\geq r_{\gamma}$.

\bibitem{Hodto} S. Hod, to be published.

\end{thebibliography}
\end{document}